# Hydrogen Peroxide Formation Rates in a PEMFC Anode and Cathode: Effect of Humidity and Temperature

**Vijay A. Sethuraman, John W. Weidner**[*]
Center for Electrochemical Engineering, Department of Chemical Engineering
University of South Carolina, 301 Main Street, Columbia, South Carolina 29208, USA

**Andrew T. Haug, Sathya Motupally, Lesia V. Protsailo**
UTC Power, 195 Governor's Highway, South Windsor, Connecticut 06074, USA

## Abstract

Hydrogen peroxide ($H_2O_2$) formation rates in a proton exchange membrane (PEM) fuel cell anode and cathode were estimated as a function of humidity and temperature by studying the oxygen reduction reaction (ORR) on a rotating ring disc electrode (RRDE). Fuel cell conditions were replicated by depositing a film of Pt/Vulcan XC-72 catalyst onto the disk and by varying the temperature, dissolved $O_2$ concentration and the acidity levels in hydrochloric acid ($HClO_4$). The $HClO_4$ acidity was correlated to ionomer water activity and hence fuel cell humidity. The $H_2O_2$ formation rates showed a linear dependence on oxygen concentration and square dependence on water activity. The $H_2O_2$ selectivity in ORR was independent of oxygen concentration but increased with decrease in water activity (*i.e.,* decreased humidity). Potential dependent activation energy for the $H_2O_2$ formation reaction was estimated from data obtained at different temperatures.

[*] – Corresponding author: Phone: (803) 777-3207; Fax: (803) 777-8265;
E-mail: weidner@engr.sc.edu (J. W. Weidner)

# Introduction

Proton exchange membrane fuel cell (PEMFC) technology, owing to its high efficiency, operational flexibility and superior modularity, has the capability to be the structural and fundamental unit of an impending hydrogen economy. Two main issues impede its progress towards commercialization are cost and durability. The US Department of Energy's (DOE) projected performance requirements[1] for the year 2010 are 5,000 hours (with 20,000 start/stops) at $45/kW for automotive stacks and upwards of 40,000 hours at $400-$750/kW for stationary power plants. In addition, current engineering requirements demand stack operation at higher temperatures (>100 °C) and low relative humidities (< 75 % RH). Elevated temperature operation offers better tolerance to CO, faster ORR kinetics, and better water and thermal management enabling easier system integration. However, elevated temperatures and the desire to operate at ambient pressures means the fuel cell needs to be operated at lower relative humidities. Since much of ionomer and membrane technologies have evolved around the water dependent perfluorinated systems such as Nafion®, both high temperature and low humidity conditions cause severe performance degradation[2-8] and remain an impediment towards achieving DOE's performance and durability targets. [2, 3, 4, 5, 6, 7, 8]

One of the mechanisms for catalyst/ionomer chemical degradation in PEMFCs involves the formation of hydroxyl and hydroperoxyl (OH• & OOH•) radicals[9, 10] caused by hydrogen peroxide ($H_2O_2$) formation on the catalyst surface via reaction 1

$$O_2 + 2H^+ + 2e^- \leftrightarrow H_2O_2 \qquad E^0 = 0.695V \quad vs.\,SHE \qquad \qquad 1$$

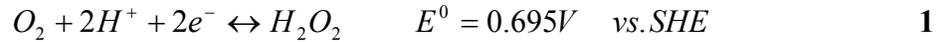

and subsequent decomposition via reactions 2 – 3. Using a novel *in situ* spin trap electron paramagnetic resonance study[11], Panchenko et al.[12] reported no evidence of OH• and OOH• radicals in the anode. They observed the presence of radicals in the cathode and near the membrane-cathode interface. Therefore, the $H_2O_2$ diffuses into the membrane and chemically breaks down to hydroxyl radicals and ions on metal ions present[13] in the membrane.

$$H_2O_2 + M^{2+} \longrightarrow M^{3+} + OH \bullet + OH^- \qquad \qquad 2$$

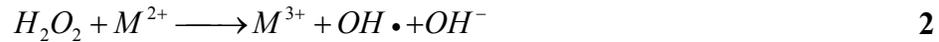

$$OH \bullet + H_2O_2 \longrightarrow OOH \bullet + H_2O \qquad \qquad 3$$

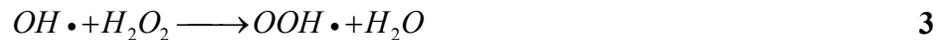

These radicals react with the perfluorosulfonic acid (PFSA) type ionomer in the electrode and the membrane to produce hydrofluoric acid (HF)[14]. The sequence is listed below

$$R_p - CF_2COOH + OH \bullet \longrightarrow R_p - CF_2 + CO_2 + H_2O \qquad \qquad 4$$

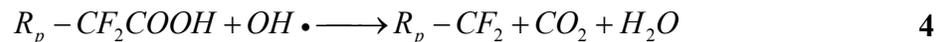

$$2R_p - CF_2 + 2OH \bullet \longrightarrow R_p - CF_2OH + R_p - COF + HF \qquad \qquad 5$$

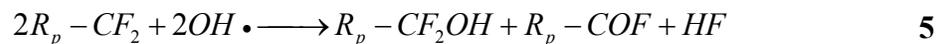

$$R_p - COF + H_2O \longrightarrow R_p - COOH + HF \qquad \qquad 6$$

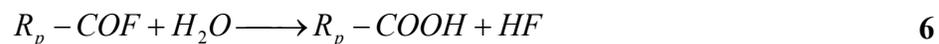

The fluoride emission rate (FER) is a measure of membrane degradation given in equations 5 and 6. Since two thirds of Nafion® is fluorine (on a mass basis), this chemical degradation results in mechanical instability in the membranes causing pinholes and eventual failure. Since this degradation is initiated by the peroxide-radical attack, understanding $H_2O_2$ kinetics at the electro-catalyst/ionomer interface at low humidities and elevated temperatures from a PEM fuel cell context is vital towards explaining the increased degradation rate observed under such conditions. Though Liu and Zuckerford[15] have reported a method for *in situ* detection of $H_2O_2$ formation, it only served as a qualitative indicator of the existence of peroxide. *In situ* quantification of peroxide kinetics is very difficult owing to its instability.[16, 17, 18, 19, 20, 21]

RRDE studies on supported Pt catalysts have been successfully used[16-21] as a technique to quantify peroxide formation and for screening oxygen reduction catalysts. Paulus et al.[16,18] reported the use of a *thin-film* RRDE method for characterizing oxygen reduction reaction (ORR) in supported high surface area catalysts and were able to quantify the amount of $H_2O_2$ produced during the oxygen reduction reaction (ORR). They decreased the film thickness and improved the ionomer-catalyst film stability at higher rotation speeds, which resulted in uniform collection efficiencies and better peroxide measurements. Antoine et al.[19] reported a weak platinum particle size effect on $H_2O_2$ production during ORR and agreed with previously reported observations that $H_2O_2$ yields were higher for potentials less than 0.4 V vs. SHE. Enayetullah et al.[20] studied ORR electrocatalysis on polycrystalline Pt microelectrode in various concentrations of Trifluoromethane sulfonic acid (TFMSA). They reported higher Tafel slopes and lower activation energies for ORR in higher concentrations of TFMSA, which was attributed to lower water activity. Murthi et al.[21] studied ORR in supported Pt and Pt alloy catalysts in 1M and 6 M TFMSA as a way to study the effect of water activity. They reported higher peroxide yields in 6 M TFMSA solution compared to a 1M solution. However none of these studies correlate the measured peroxide yields and selectivity to peroxide formation rates in a PEM fuel cell as a function of cell operating conditions.

Therefore, the objective of this investigation is to predict $H_2O_2$ formation rates in a PEM fuel cell. This was accomplished by measuring $H_2O_2$ formation rates at rotating ring disc electrode (RRDE). Fuel cell conditions were replicated by depositing a film of Pt/Vulcan XC-72 catalyst onto the disk and by varying the temperature, dissolved $O_2$ concentration and the acidity levels in $HClO_4$. The $HClO_4$ acidity was correlated to ionomer water activity and hence fuel cell humidity. Peroxide formation rate in the anode was predicted using oxygen permeability measured across Nafion® 112 membrane and the fraction of oxygen that reduces to H2O2. Peroxide formation rate in the cathode was predicted using the rate constants measured from RRDE experiments and the local concentrations of oxygen and protons.

## Experimental

*Rotating Ring Disc Electrode (RRDE)* – For the RRDE studies, commercially available Pt/Vulcan catalyst (20% Pt on Vulcan XC-72R carbon, Johnson Matthey Inc., PA) was used. Catalyst coated glassy carbon electrodes were prepared as described by Schmidt et al.[22]. Aqueous suspensions of 1 mg catalyst ml$^{-1}$ were obtained by pulse-sonicating 20 mg Pt/Vulcan catalyst with 20 ml triple-distilled, ultrapure water (Millipore Corporation) in an ice bath (70% duty cycle, 60W, 15 minutes). Sonication was done using a Braun-Sonic U Type 853973/1 sonicator. A glassy carbon disc served as the substrate for the supported catalyst and was polished to a mirror finish (0.05 μm deagglomerated alumina, Buehler®) prior to catalyst coating. An aliquot of 14 μl catalyst suspension was pipetted onto the carbon substrate, which corresponded to a Pt loading of ~14.1 μg Pt cm$^{-2}$. After evaporation of water for 30 minutes in $N_2$ atmosphere (15 in-Hg, vacuum), 14 μl of diluted Nafion solution (5% aqueous solution, 1100 EW; Solution Technology Inc., Mendenhall, PA) was pipetted on the electrode surface and further evaporated for 30 minutes in $N_2$ atmosphere (15 in-Hg, vacuum). Nafion® was used to adhere the Pt/Vulcan particles onto the glassy carbon electrode (the ratio of $H_2O$/Nafion® solution used was ca. 100/1). Previous work by Paulus et al. indicate that this procedure yielded a Nafion® film thickness of ca. 0.1 μm and that the utilization of the Pt/Vulcan catalyst (based on H-adsorption charge) on the electrode with this film was ~100%.

The catalyst-Nafion® coated electrode was immersed in deaerated (UHP Nitrogen, Praxair) Perchloric acid ($HClO_4$, 70%, ULTREX II® Ultrapure Reagent Grade, J. T. Baker) of varying concentrations for further synchronized chrono-amperometric and potentiodynamic experiments. Though a variety of supporting electrolytes are reported in the literature, anion adsorption on Pt is minimal for only a few electrolytes[23] (e.g., Trifluoromethane sulfonic acid (TFMSA) and $HClO_4$). In addition, the ultrapure reagent grade $HClO_4$ used in this study is free of ionic impurities; especially since $Cl^-$ ions, even in trace amounts (i.e. 1 ppm), is shown to drastically change both the activity and the reaction pathway of ORR on Pt catalysts.[23, 24, 25] All RRDE experiments were performed at atmospheric pressure and all solutions were prepared from ultrapure water (Millipore Inc., 18.2 MΩcm).

The electrochemical measurements were conducted in a standard electrochemical cell (RDE Cell®, Pine Instrument Company, NC) immersed in a custom-made jacketed vessel, temperature of which was controlled by a refrigerated/heating circulator (Julabo Labortechnik GMBH). A ring-disk electrode setup with a bi-potentiostat (Bi-Stat®, Princeton Applied Research Inc., TN) in conjunction with rotation-control equipment (Pine Instrument Company, NC). EC-Lab® software (version 8.60, Bio-logic Science Instruments, France) was used to control the bi-potentiostat. The Pt ring electrode was held at 1.2 V vs. SHE where the oxidation of peroxide is diffusion limited. The catalyst coated glassy carbon disc electrode (5 mm diameter, 0.1966 cm$^2$ area, DT21 Series, Pine Instrument Company, NC) was scanned between 0 – 1.2 V vs. SHE to characterize $H_2O_2$ formation within the potential range relevant to fuel cell operating conditions. Potentials were determined using a mercury-mercurous sulfate (Hg/$Hg_2SO_4$) reference electrode. All potentials in this study, however, refer to that of the standard hydrogen electrode (SHE). A high-surface area Pt cylindrical-mesh (5 mm diameter, 50 mm length) attached to a Pt wire (0.5 mm thick, 5 mm length) was used as the counter electrode.

*Effect of oxygen concentration* – The effect of oxygen concentration on ORR and $H_2O_2$ formation kinetics was studied by varying the concentration of oxygen in the solution. The following three gases were used: oxygen (UHP grade, Praxair), Air (Industrial, Praxair) and 10.01% oxygen in nitrogen (Airgas). A gas flow meter (0-500 ml, Dwyer Instruments Inc., IN) was used to control the flow of the gas feed at ~100 ml min$^{-1}$ into the electrolyte. The electrochemical cell was sealed during the experiments to keep air from affecting the concentration of dissolved oxygen in the electrolyte. The concentration of dissolved oxygen in the electrolyte was estimated using the solubility values for oxygen in pure liquid water at 25 ˚C and 101 kPa[26].

*Effect of pH* – The effect of proton concentration on ORR and $H_2O_2$ formation kinetics was studied by varying the acidity of $HClO_4$ in the 2.0 – 0.1 M concentration window (~-0.301 – 1 pH, assuming $K_a$ >>1 for $HClO_4$). Between solution changes, the electrochemical cell and its components were washed and boiled in DI water for 5 hours to ensure accurate pH levels. The catalyst-Nafion® coated electrode was also cleaned in a sonicator before every experiment with triple distilled ultrapure water.

*Collection efficiency* – Standard procedure[27] for the determination of collection efficiency of a ring-disc electrode was followed. The electrodes were prepared as described above. The experiment was carried out in an electrochemical cell in deaerated (UHP Nitrogen, Praxair) 0.1 M $H_2SO_4$ (96.5%, J. T. Baker) with 10 mmol l$^{-1}$ $K_3Fe(CN)_6$ (99.7%, J. T. Baker. The disk electrode was swept at 1 mV s$^{-1}$ [vs. SHE] while the Pt ring was held at a constant potential of 1.2 V [vs. SHE]. At this ring potential, the oxidation of $[Fe(CN)_6]^{4-}$, produced at the disk electrode, to $[Fe(CN)_6]^{3-}$, proceeds under pure diffusion control. The collection efficiency was determined as *N = $I_{ring}/I_{disk}$ = 0.20*, which was independent of disk potential and consistent with the theoretical collection efficiency provided by the manufacturer of the ring-disc electrode[28].

## Theory

The two-electron transfer reaction of $O_2$ reduction to $H_2O_2$, captured by the Pt ring, was analyzed in this work. At the ring, the $H_2O_2$ produced at the disk is oxidized via the reverse of reaction 1. The fraction of $H_2O_2$ formation, $\chi_{H_2O_2}$, can be determined from the collection efficiency, ring and disk currents by the expression,

$$\chi_{H_2O_2} = \frac{2I_{ring}/N}{I_{disk} + I_{ring}/N} \qquad 7$$

The measured current density *j* corresponding to $H_2O_2$ formation on a film covered RDE for the first-order ORR kinetics was previously reported to take the following expression[29], in terms of kinetic and mass-transport dependent currents,

$$\frac{1}{j} = \frac{1}{j_{kin}} + \frac{\delta_f}{nFD^f_{O_2}C^f_{O_2}} + \frac{1}{j_D} \qquad 8$$

Where j is

$$j = \frac{I_{ring}}{NA} \qquad 9$$

$j_{kin}$ is the current density in the absence of mass transfer effects and $j_D$ is the diffusion current given by the Levich equation.

$$j_D = 0.62 nF D_{O_2}^{*\,2/3} C_{O_2}^* v^{-1/6} \omega^{1/2} \qquad 10$$

The concentration of $O_2$ in the solution was calculated from the partial pressure of $O_2$ in the inlet gas and $O_2$ solubility data for pure liquid water at corresponding temperature and 101 kPa[26]. The difference in $O_2$ solubility in pure liquid water and in $HClO_4$ (up to 2M) was assumed to be negligible. Combining equations 8 and 10 and solving for $j_{kin}$ gives,

$$j_{kin} = \frac{j n F D_{O_2}^f C_{O_2}^f D_{O_2}^{*\,2/3} C_{O_2}^* \omega^{1/2}}{n F D_{O_2}^f C_{O_2}^f D_{O_2}^{*\,2/3} C_{O_2}^* \omega^{1/2} - \delta_f j D_{O_2}^{*\,2/3} C_{O_2}^* \omega^{1/2} - 1.6 v^{1/6} j D_{O_2}^f C_{O_2}^f} \qquad 11$$

The purely kinetic portion of the $H_2O_2$ formation rate is

$$R_{H_2O_2} = \frac{j_{kin}}{2F} = k_f \left(C_{O_2}\right)^a \left(C_{H^+}\right)^b \qquad 12$$

Where,

$$k_f = k_f^0 \exp\left[\frac{\alpha F \eta}{R T^0}\right] \qquad 13$$

In equation 12, '$a$' and '$b$' are reaction orders with respect to $O_2$ and $H^+$ respectively. Only the forward rate term is used in equation 12 because at 0.6V vs. SHE and below, the rate of oxidation of $H_2O_2$ (the reverse reaction) is negligible. The kinetic rate constant $k_f$ was estimated for different potentials by plotting $H_2O_2$ production rate as a function of oxygen concentration for various potentials. Since the electrode reaction rate was earlier shown by Damjanovic and Hudson[30] to be faster on an oxide-free Pt surface than on an oxide-covered surface, both the forward and the reverse scans were used to estimate the reaction rate constant. The potential dependence of this rate constant is given in equation 13.

The activation energies for hydrogen peroxide formation reaction were evaluated by using the Arrhenius equation[31] shown below,

$$k_f^0 = k_{f,0}^0 \exp\left[\frac{E_a}{RT}\right] \qquad 14$$

The activation energies for H$_2$O$_2$ formation on supported Pt catalysts were compared to the computationally estimated activation energies reported in the literature. For example, using density functional theory (DFT), Anderson and Albu[32], Sidik and Anderson[33] and Wang and Balbuena[34] have reported activation energies for H$_2$O$_2$ formation on Pt$_1$, Pt$_2$ and Pt$_3$ sites respectively.

**Results and Discussion**

*H$_2$O$_2$ kinetic studies using RRDE* – Figure 1a shows polarization curves for the oxygen reduction reaction and Figure 1b shows ring currents corresponding to H$_2$O$_2$ oxidation. It should be noted that hydrogen evolution starts in the neighborhood of ca. 25-50 mV vs. SHE. Hence, both ring and disc currents in the 25-50 mV region include a fractional contribution due to hydrogen evolution [$2H^+ + 2e^- \longrightarrow H_2$] and oxidation [$H_2 \longrightarrow 2H^+ + 2e^-$] respectively. This is noted by the decrease in the disk currents and an increase in the ring currents in the 25 – 50 mV potential region. Data obtained below 25 mV was not used in this analysis. Figure 1c shows the fraction of peroxide produced in the ORR at the disk, as captured by the ring and shows no dependence on oxygen concentration.

Figure 2 shows the purely kinetic portion of H$_2$O$_2$ formation rates obtained from equation 12 as a function of oxygen concentration in the 2M HClO$_4$ for different voltages. The data shows a linear dependence of the oxygen concentration at all potentials (i.e., a = 1). In this figure, overpotential of 0.670 V represents a potential of 0.025 V vs. SHE because the equilibrium potential for H$_2$O$_2$ formation is 0.695 V. Four representative overpotentials were chosen for this plot. The anode experiences the highest overpotential for peroxide formation during fuel cell operation. The cathode potential is above the H$_2$O$_2$ equilibrium potential at open circuit but experiences a significant drop during load conditions, and can go negative of H$_2$O$_2$ equilibrium potential.

Figure 3 shows the potential dependence of the rate constants estimated from equation 12. The 25 ºC data between η values 0-0.3V and 0.3-0.65V was fit with two separate linear equations in Figure 3. The respective intercepts represent $k_f^0$ (equation 13) and are independent of $C_{H^+}$. These values, as a function of T, were used to obtain activation energy.

Figure 4 shows the ring currents and the fraction of H$_2$O$_2$ formed with different acidities. Perchloric acid systems in the 0.1 and 2.0 M concentration range equilibrated with pure O$_2$ were used to study the effect of proton concentration. Figure 4 shows an increased rate of H$_2$O$_2$ formation with increasing proton concentration. Since the disc currents were similar for all acid concentrations, the increased ring currents meant that selectivity towards peroxide formation was a function of proton concentration.

Figure 5 shows the dependence of H$_2$O$_2$ formation rate on proton concentration. This is consistent with Murthi et al.'s RRDE results in 1 and 6 M TFMSA. The points are measurements and lines are predictions according to equation 12. The reaction order with respect to $H^+$ in the $H_2O_2$ formation reaction was found to best fit the data for b = 2. The reason for the change in selectivity shown by the Pt catalyst in favor of the two-electron transfer at higher acid concentrations (or lower water activities) is beyond the scope of this article. There is a possibility that the adsorption of $ClO_4^-$ anions could influence the reaction pathway in ORR on Pt catalysts similar to that of $Cl^-$ anions as previously reported by Schmidt et al.[23], Stamenkovic et al.[25], and Markovic and Ross.[24] However, the bond strength of $ClO_4^-$ adsorption on Pt is much weaker than $Cl^-$ and $SO_4^{2-}$ and its influence on the reaction pathway of ORR is also very minimal.[23, 25]

The $H_2O_2$ formation rates measured as a function of water activity, potential and temperature using RRDE experiments was used to predict $H_2O_2$ formation rates at the anode and cathode of PEM fuel cell. Peroxide formation rate at the anode was predicted using oxygen permeability from the cathode and $\chi_{H_2O_2}$. Peroxide formation rate at the cathode was predicted via equation 12, i.e. as a product of the rate constant and the local reactant concentrations. Peroxide formation at the cathode occurs only for fuel cells operating under considerable load (i.e., high cell current) such that the local potential goes negative relative to the equilibrium potential for peroxide formation. For estimation of peroxide rates, local potential at the cathode was taken to be 0.6 V (i.e., η = 0.095 V).

Nafion® is a super-acid catalyst and hence the local acidity at the catalyst-membrane interfaces was calculated from the local water content and the fixed number of sulfonic acid groups. The water sorption properties of Nafion® as a function of temperature and water activity had been studied by several laboratories[33-37]. [35, 36, 37, 38, 39]

Using a novel tapered element oscillating microbalance (TEOM) technique, Jalani et al.[37] measured water uptake in Nafion as a function of water activity in vapor phase between 30 °C and 110 °C and reported that the water uptake increased with temperature and was highest at 110 °C. The difference in water uptake between 30 °C and 110 °C is negligible for lower water activities ($a_w$ < 0.7). This is reported by Jalani et al (experimental) and discussed in detail by Motupally et al. (simulations)[40]. For this work, the absorption isotherm of Nafion® 117 membranes measured at 30 °C by Zawodzinski et al.[38] was used. Between water activity values of 0 and 1, the experimentally measured absorption isotherm was fit to the following polynomial[41],

$$\lambda = 0.043 + 17.81[a_w] - 39.85[a_w]^2 + 36.0[a_w]^3 \qquad \textbf{15}$$

In this equation, λ represents the number of water molecules per sulphonic acid group in the polymer and $a_w$ represents the activity of water, which is the effective mole fraction of water given by $p_0/p^*$, where $p^*$ is the vapor pressure of water, in bar. $p^*$ was calculated from the Antoine correlation,

$$\ln p^* = A_1 - \frac{B_1}{T + C_1} \qquad \textbf{16}$$

The constants are $A_1 = 11.6832$, $B_1 = 3816.44$, $C_1 = -46.13$[42]. Inside a fuel cell, this water activity is essentially the equilibrium relative humidity expressed as a fraction. The concentrations of $H_2O$ and $H^+$ in the polymer are respectively expressed as,

$$C_{H_2O} = \frac{\rho \lambda}{EW} \qquad 17$$

$$C_{H^+} = \frac{C_{H_2O}}{\lambda} \qquad 18$$

In these equations, EW is the equivalent weight of the polymer (taken to be 1100) and $\rho$ is the humidity-dependent density of the polymer given by,

$$\rho = \frac{1.98 + 0.0324\lambda}{1 + 0.0648\lambda} \qquad 19$$

It was assumed that all sulphonic acid groups exist in a completely dissociated form. Figure 6 shows the variation of $\lambda$ and pH of Nafion® as a function of water activity. Even at vapor-saturated conditions [$\lambda = 14$], the pH of Nafion® is below 0. This trend (not shown) is seen for $HClO_4$, also a strong acid. The acid was assumed to be completely dissociated i.e., Ka >>1. This approach in relating MEA acidity to water activity and hence to the humidity of the incoming gases facilitates in computing peroxide rates inside a fuel cell. Quantitatively, the measured peroxide rates via the RRDE experiments at a particular oxygen concentration, pH value and temperature should equal the peroxide rates inside the fuel cell at same pH value and temperature. Since rate constants were measured at 25 °C using RRDE, they had to be estimated at higher temperatures in order to be used to calculate peroxide formation rates in a fuel cell. Activation energies for $H_2O_2$ formation on supported Pt catalysts were estimated from kinetic currents obtained at 15 °C, 25 °C, 35 °C and 45 °C. Figure 7 shows the Arrhenius plot of the logarithmic pre-exponential term $k_f^0$ versus the inverse of temperature for the two overpotential regions (0-0.3V and 0.3-0.65V). The activation energies were obtained from the slope ($E_a/R$) of a linear fit according to equation 14. This procedure is analogous to those described by Neyerlin et al.[43] and Bard and Faulkner[44].

*Oxygen flux across Nafion® at higher temperatures* – Oxygen permeability through Nafion® depend greatly on the water content of the membrane. It has been shown by Sakai et al.[45] that $O_2$ diffusion rates in a completely dry Nafion® membrane has values similar to that in PTFE and approaches the limit of liquid water with increasing water content. Figure 8 shows experimentally measured $O_2$ permeability, corrected for an $O_2$ feed at 101 kPa, across a Nafion® 112 membrane as a function of temperature and relative humidity. These permeation rates were estimated using electrochemical monitoring technique (EMT) and are comparable to those estimated by gas chromatography (GC) method[46]. Between 25% and 100% relative humidity of the feed gas, the permeabilities differ by as much as an order of magnitude. Permeability for other temperatures and water contents were estimated by the following equation which was derived by fitting the measured permeability values,

$$P_{O_2}^m = \left(1.002 \times 10^{-14} - 9.985 \times 10^{-15} a_w\right) \exp\left[\left(0.0127 + 2.3467 \times 10^{-2} a_w\right) T\right] \qquad 20$$

Oxygen solubility at the membrane-cathode catalyst layer interface, $C_{O_2}^c$, was estimated using the following relation,

$$C_{O_2}^c = \frac{P_{O_2}^m}{D_{O_2}^m} \qquad 21$$

$D_{O_2}^m$ values for different temperatures and relative humidities were obtained from Sakai et al.'s work[45] and was fit to the following expression,

$$D_{O_2}^m = 9.78 \times 10^{-8} + 3.5 \times 10^{-9} T + 10^{-4} a_w \qquad 22$$

Figure 9 shows the peroxide rates at the cathode-membrane interface when the local potential at the cathode is 0.6 V and the gas feed is pure oxygen at 1 atm. Though the amount of peroxide formed at the cathode is only of the order of few micromoles at fully humidified conditions, the relative difference between dry and fully humidified conditions at 95 ºC is significant.

The potential profile across the membrane, measured *in situ* by Liu and Zuckerbrod[15] [figure 19] and modeled by Burlatsky et al.[47] at open circuit conditions, indicate that the potential at the anode-membrane interface is ~ 0 V. For the purpose of calculating H$_2$O$_2$ rates at the anode/membrane interface, a potential of ~0 V (i.e., η = 0.695 V) was assumed to exist at the interface.

The oxygen flux across the membrane from the cathode to the anode is,

$$F_{O_2} = \frac{D_{O_2}^m}{\delta}\left(C_{O_2}^c - C_{O_2}^a\right) \qquad 23$$

The concentration of oxygen at the anode-membrane interface approaches zero, assuming all of the oxygen crossing over the membrane to the anode side is reduced to water or reacts chemically with hydrogen.

$$R_{H_2O_2}^a = \chi_{H_2O_2} \frac{P_{O_2}^m}{\delta} \qquad 24$$

While the fraction of oxygen that is reduced to peroxide is a strong function of water activity, it is not a function of oxygen concentration (see Figure 1c). An expression for $\chi_{H_2O_2}$ versus $C_{H^+}$ was obtained from measured values at room temperature,

$$\chi_{H_2O_2} = 0.2081 - 0.1208(a_w) - 0.072(a_w)^2 - 2.132 \times 10^{-14}(a_w)^3 \qquad 25$$

Figure 10 shows the $H_2O_2$ formation rates at the anode-membrane interface. This goes through a peak because oxygen permeability decreases with decreasing water activity whereas $H_2O_2$ selectivity increases with decrease in water activity.

These anode and cathode $H_2O_2$ formation rates cannot be directly correlated with the fluoride emission rates because there are several intermediate reactions between $H_2O_2$ formation and actual membrane degradation. Though additional mechanisms involving direct radical formation on Pt followed by their chemical attack on the membrane were suggested[47], they need further experimental validation and verification. However quantification of $H_2O_2$ formation rates described in this work is important in the mechanistic understanding of membrane degradation. This quantification would help in validating a durability mechanism especially at elevated temperatures and low relative humidities.

## Conclusion

$H_2O_2$ formation rates in a PEM fuel cell anode and cathode were estimated by studying the ORR kinetics on a ring disc electrode. Fuel cell conditions were replicated by depositing a film of Pt/Vulcan XC-72 catalyst onto the disk and by varying the temperature, dissolved $O_2$ concentration and the acidity levels in $HClO_4$. The $HClO_4$ acidity was correlated to ionomer water activity and hence fuel cell humidity. $H_2O_2$ formation rates showed a linear dependence on oxygen concentration and a square law dependence on water activity. The $H_2O_2$ selectivity in ORR was independent of oxygen concentration but increased with decrease in water activity (i.e., decreased humidity). Potential dependent activation energy for the $H_2O_2$ formation reaction was estimated from data obtained at different temperatures. Anode peroxide formation rates are proportional to the oxygen flux from the cathode and was estimated to be three orders of magnitude lower than cathode formation rates for a cell operating under load conditions (i.e., V ≤ 0.6V).

# Tables

**Table 1. Parameters used in the analysis of measured current at the Pt ring[a].**

| Parameter | Value | Comments |
|---|---|---|
| $a$ | 1 | Measured |
| $A$ | 0.164025 cm$^2$ | Ref. 28 |
| $b$ | 2 | Measured |
| $C^*_{O_2}$ | 1.274 mol cm$^{-3}$ | Ref. 26[a] |
| $D^*_{O_2}$ | 2.2 x 10$^{-6}$ cm$^2$ s$^{-1}$ | Ref. 48 |
| $E^0$ | 0.695 V vs. SHE | |
| $EW$ | 1100 g mol$^{-1}$ | Ref. 49 |
| $F$ | 96485 C mol equiv.$^{-1}$ | Ref. 50 |
| $N$ | 0.2 | Measured |
| $R$ | 8.314 J mol$^{-1}$ K$^{-1}$ | Ref. 50 |
| $T^0$ | 298 K | Measured |
| $\alpha$ | 0.5 | Assumed |
| $\delta_f$ | 10$^{-5}$ cm | Ref. 16 |
| $\rho$ | 1 g cm$^{-3}$ | Ref. 50 |
| $\upsilon$ | 0.009 cm$^2$ s$^{-1}$ | Estimated |
| $\omega$ | 2500 s$^{-1}$ | Measured |

[a] The mole fraction solubility $X_1$ of oxygen in water is given as $\ln X_1 = A_2 + \frac{B_2}{T^*} + C_2 \ln T^*$, where, $T^* = \frac{T}{100} K$. A$_2$ = -66.7354, B$_2$ = 87.4755 and C$_2$ = 24.4526.

# List of symbols

| | |
|---|---|
| $a$ | reaction-order with respect to $O_2$ in the $H_2O_2$ formation reaction |
| $a_w$ | water activity |
| A | disk area, cm$^2$ |
| $b$ | reaction-order with respect to $H^+$ in the $H_2O_2$ formation reaction |
| $C_{H^+}$ | proton concentration, mol cm$^{-3}$ |
| $C_{H_2O}$ | water concentration in the membrane, mol cm$^{-3}$ |
| $C_{O_2}^*$ | oxygen concentration in the bulk of the electrolyte, mol cm$^{-3}$ |
| $C_{O_2}^f$ | oxygen concentration in Nafion® film, mol cm$^{-3}$ |
| $C_{O_2}^a$ | oxygen concentration in Nafion® 112 membrane-anode catalyst layer interface, mol cm$^{-3}$ |
| $C_{O_2}^c$ | oxygen concentration in Nafion® 112 membrane-cathode catalyst layer interface, mol cm$^{-3}$ |
| $D_{O_2}^*$ | oxygen diffusion coefficient in the electrolyte, cm$^2$ s$^{-1}$ |
| $D_{O_2}^f$ | oxygen diffusion coefficient in Nafion® film, cm$^2$ s$^{-1}$ |
| $D_{O_2}^m$ | diffusion coefficient of $O_2$ in Nafion® 112 membrane, cm$^2$ s$^{-1}$ |
| $E^0$ | equilibrium potential, 0.695 V vs. SHE at 25 ˚C and 101 kPa |
| $E_a^*$ | activation energy for $H_2O_2$ formation, J mol$^{-1}$ |
| $E_{app}$ | applied potential, V vs. SHE |
| EW | equivalent weight of Nafion® polymer, 1100 g equiv$^{-1}$ |
| F | Faraday constant, 96485 C mol$^{-1}$ |
| $I_{ring}$ | ring current, mA |
| $I_{disk}$ | disk current, mA |
| $j$ | total peroxide current density, mA cm$^{-2}$ |
| $j_{disk}$ | disk current density, mA cm$^{-2}$ |
| $j_D$ | diffusion-limited current density, mA cm$^{-2}$ |
| $j_{kin}$ | kinetic current density, mA cm$^{-2}$ |
| $k_b$ | rate constant for $H_2O_2$ electro-oxidation, s$^{-1}$ |
| $k_f$ | rate constant for $H_2O_2$ formation, mol$^2$ cm$^{-5}$ s$^{-1}$ |
| N | collection efficiency |
| $n$ | number of electrons transferred per $O_2$ molecule in $H_2O_2$ formation, 2 |
| $P_{O_2}^m$ | permeability of $O_2$ in Nafion® 112 membrane, mol cm$^{-1}$ s$^{-1}$ |
| R | universal gas constant, 8.314 J mol$^{-1}$ K$^{-1}$ |
| T | temperature, K |
| $t$ | time, s |

**Greek**

| | |
|---|---|
| $\alpha$ | transfer co-efficient |
| $\delta$ | Pt/C electrode thickness, cm |
| $\delta_f$ | Nafion® film thickness, cm |
| $\rho$ | density of Nafion®, g cm$^{-3}$ |
| $v$ | kinematic viscosity, cm$^2$ s$^{-1}$ |
| $\eta$ | overpotential, V vs. SHE |
| $\lambda$ | moles of water per sulphonic acid group in Nafion® |
| $\chi_{H_2O_2}$ | fraction of O$_2$ reducing to H$_2$O$_2$ |
| $\omega$ | electrode rotation rate, s$^{-1}$ |

**Superscript**

| | |
|---|---|
| 0 | standard state or equilibrium |
| a | anode |
| c | cathode |

**Subscript**

| | |
|---|---|
| *b* | backward reaction |
| *D* | diffusion |
| *disk* | Pt/Nafion® coated disc electrode |
| *f* | Nafion® film or forward reaction |
| *kin* | kinetic |
| *ring* | Pt ring electrode |

## Acknowledgements

The US Department of Energy supported this work under contract number DOE-DE-FC36-02AL, for which the authors are grateful. The authors thank Mr. Michael Fortin for his help in experimental setup for oxygen crossover measurements.

# Figure captions

Figure 1: (a) Polarization curves for the oxygen reduction reaction on a Pt/Vulcan XC-72R (14.1 μg Pt cm$^{-2}$) thin-film RRDE [2500 rpm] in 2.0 M HClO$_4$ solution [pH = -0.3] bubbled with 10% O$_2$ [○], Air [▲] and O$_2$ [●].

(b) Ring currents corresponding to H$_2$O$_2$ oxidation [E$_{ring}$ = 1.2 V vs. SHE].

(c) % H$_2$O$_2$, given by equation 11, formed during the oxygen reduction reaction on Pt/Vulcan XC-72R in 2.0 M HClO$_4$ solution [pH = -0.3] bubbled with 10% O$_2$ [○], Air [▲] and O$_2$ [△] at 25 °C. E$_{ring}$ = 1.2 V vs. SHE, 1mV/s, 2500 rpm.

Figure 2: H$_2$O$_2$ formation rates [mol cm$^{-2}$ s$^{-1}$] on Pt/Vulcan XC-72R in 2.0 M HClO$_4$ solution [pH = -0.3] as a function of dissolved oxygen concentration [mol L$^{-1}$] for the following four overpotentials: 0.695 V [○], 0.495 V [△], 0.295 V [□] and 0.095 V [◇]. The symbols represent data and the lines represent linear fits with zero intercepts.

Figure 3: Electrochemical rate constant for H$_2$O$_2$ formation, $k_f$, as a function of overpotential, η = E$_{app}$ – E$^0$, E$^0$ = 0.695 V vs. SHE. The data between η values 0-0.3V and 0.3-0.65V was fit with two separate linear equations. The value of α = 0.05, T = 25 ºC. Data was obtained in 2M HClO$_4$ bubbled with pure O$_2$ at 1 atm.

Figure 4: (a) Rate and (b) % of H$_2$O$_2$ formed during the oxygen reduction reaction on Pt/Vulcan XC-72R in 0.1 M [○], 1.0 M [▲], 1.8 M [◇] and 2.0 M [●] HClO$_4$ solution [pH = -0.3] bubbled with O$_2$ at 25 °C. E$_{ring}$ = 1.2 V vs. SHE, 1mV/s, 2500 rpm.

Figure 5: Mass transport corrected H$_2$O$_2$ formation rates [mol cm$^{-2}$ s$^{-1}$] on Pt/Vulcan XC-72R in HClO$_4$ solution as a function of acidity [M] for the following three overpotentials: 0.695 V [○], 0.345 V [♦] and 0.095 V [△]. The lines are predictions according to Equation **13**. The reaction order with respect to H$^+$ in the H$_2$O$_2$ formation reaction, $b$ = 2.

Figure 6: (a) *pH* (-----) and *λ* (——) vs. water activity, $a_w$, plots for Nafion®. λ is the amount of water per sulphonic acid group [mol basis] and a$_w$ is the effective mole fraction of water and is equal to the equilibrium relative humidity expressed as a fraction.

Figure 7: Arrhenius plots for H$_2$O$_2$ formation reaction on Pt/Vulcan XC-72R in 2M HClO$_4$ at overpotential of 0.195 V (-○-) and 0.695 V (-●-) vs. NHE.

Figure 8: Oxygen permeability [mol cm$^{-1}$ s$^{-1}$] in Nafion® 112 measured by electrochemical monitoring technique as a function of temperature for two different relative humidities, 25% and 100% RH. These rates were normalized to a 101 kPa O$_2$ feed.

Figure 9: Rates of H$_2$O$_2$ formation [mol cm$^{-2}$ s$^{-1}$] in the cathode side of a PEM fuel cell for different relative humidities and temperatures. Local potential at the cathode was assumed to be ~0.6 V vs. SHE, which translates to an overpotential of 0.095 V for H$_2$O$_2$ formation.

Figure 10: Rates of H$_2$O$_2$ formation [mol cm$^{-2}$ s$^{-1}$] in the anode side of a PEM fuel cell for different relative humidities and temperatures. Local potential at the anode was assumed to be ~0 V vs. SHE, which translates to an overpotential of 0.695 V for $H_2O_2$ formation.

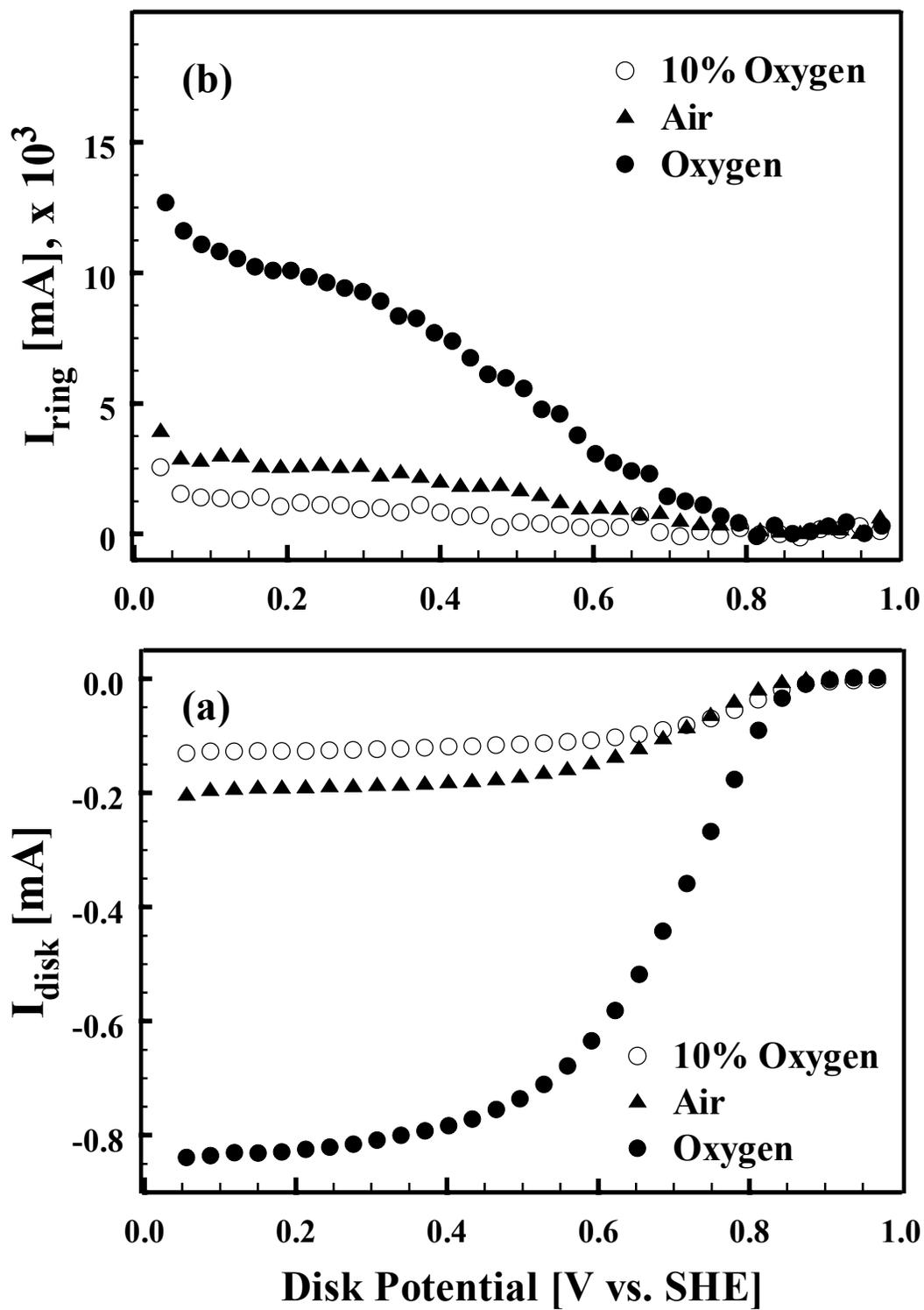

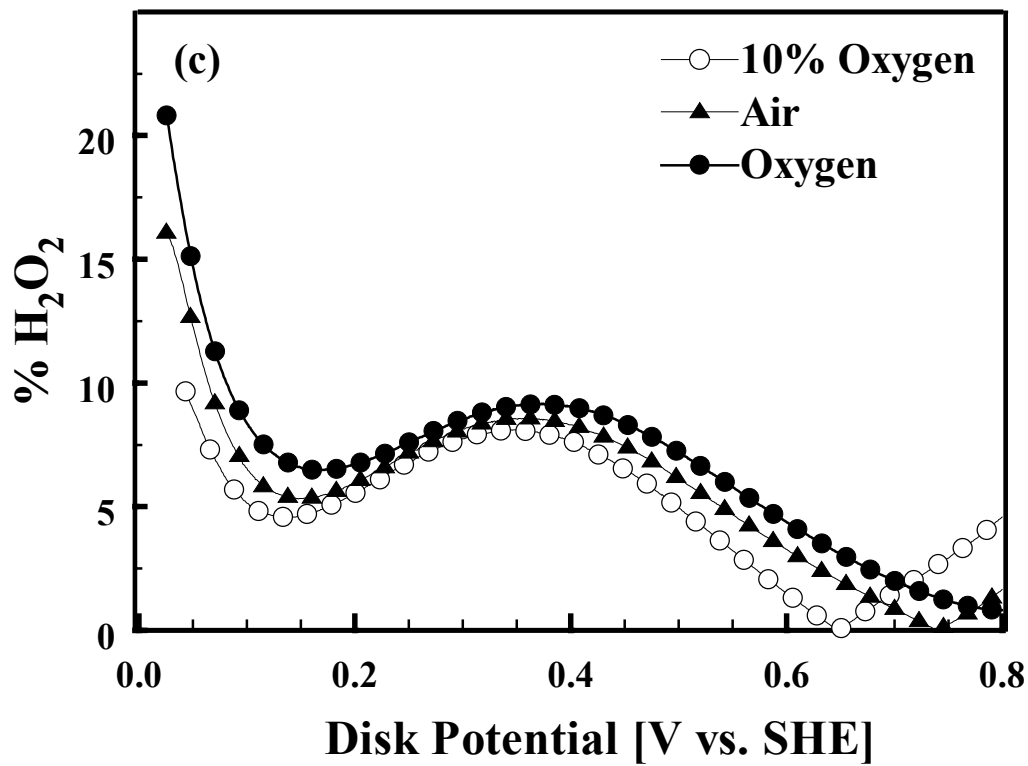

Figure 1: (a) Polarization curves for the oxygen reduction reaction on a Pt/Vulcan XC-72R (14.1 µg Pt cm$^{-2}$) thin-film RRDE [2500 rpm] in 2.0 M HClO$_4$ solution [pH = -0.3] bubbled with 10% O$_2$ [○], Air [▲] and O$_2$ [●].

(b) Ring currents corresponding to H$_2$O$_2$ oxidation [E$_{ring}$ = 1.2 V vs. SHE].

(c): % H$_2$O$_2$, given by equation 11, formed during the oxygen reduction reaction on Pt/Vulcan XC-72R in 2.0 M HClO$_4$ solution [pH = -0.3] bubbled with 10% O$_2$ [○], Air [▲] and O$_2$ [●] at 25 °C. E$_{ring}$ = 1.2 V vs. SHE, 1mV/s, 2500 rpm.

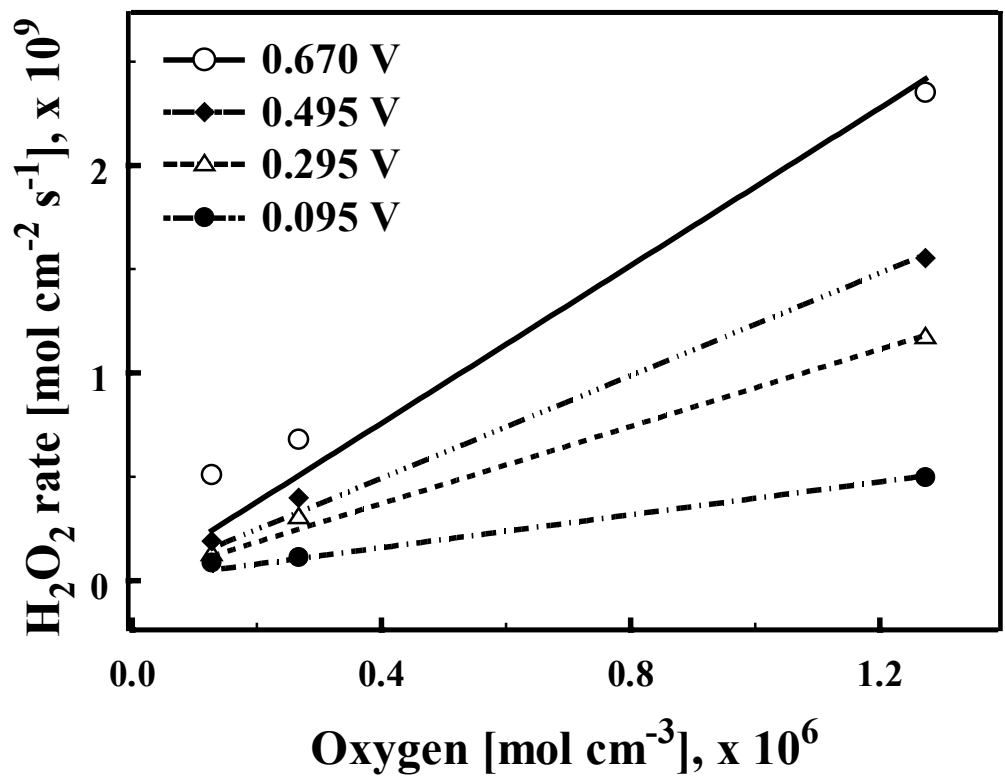

Figure 2: $H_2O_2$ formation rates [mol cm$^{-2}$ s$^{-1}$] on Pt/Vulcan XC-72R in 2.0 M HClO$_4$ solution [pH = -0.3] as a function of dissolved oxygen concentration [mol L$^{-1}$] for the following four overpotentials: 0.670 V [○], 0.495 V [◆], 0.295 V [▲] and 0.095 V [●]. The symbols represent data and the lines represent linear fits with zero intercepts.

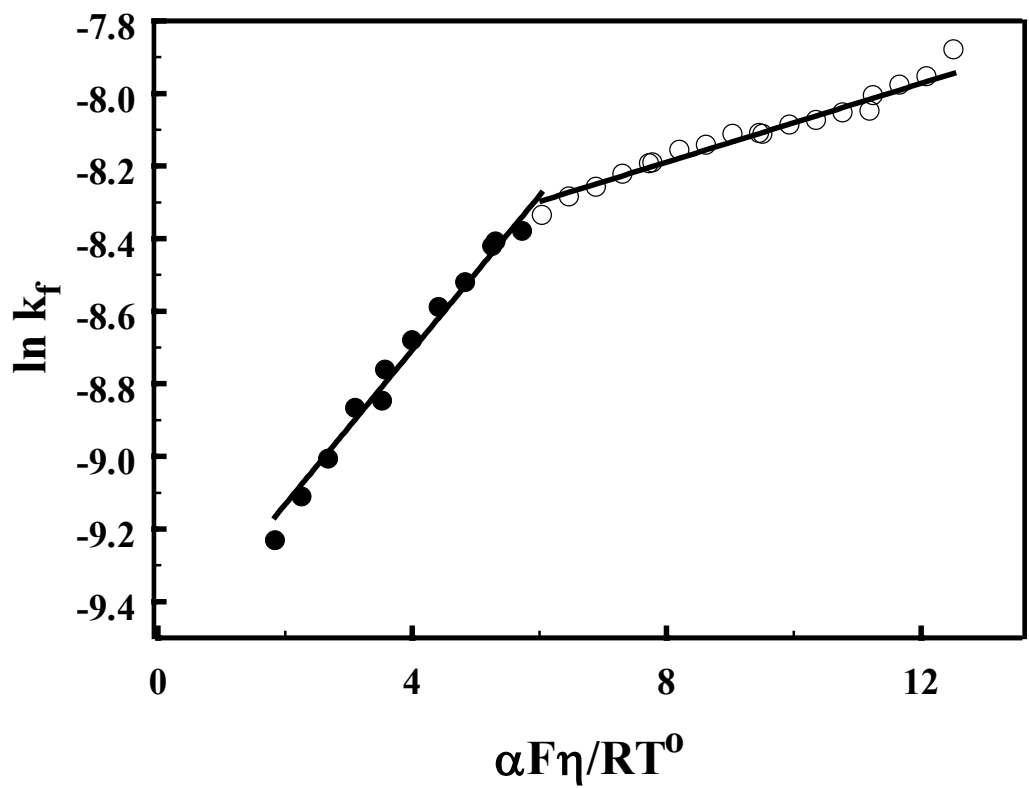

Figure 3: Electrochemical rate constant for $H_2O_2$ formation, $k_f$, as a function of overpotential, $\eta = E_{app} - E^0$, $E^0 = 0.695$ V vs. SHE. The data between $\eta$ values 0-0.3V and 0.3-0.65V was fit with two separate linear equations. The value of $\alpha = 0.05$, $T^0 = 25$ °C.

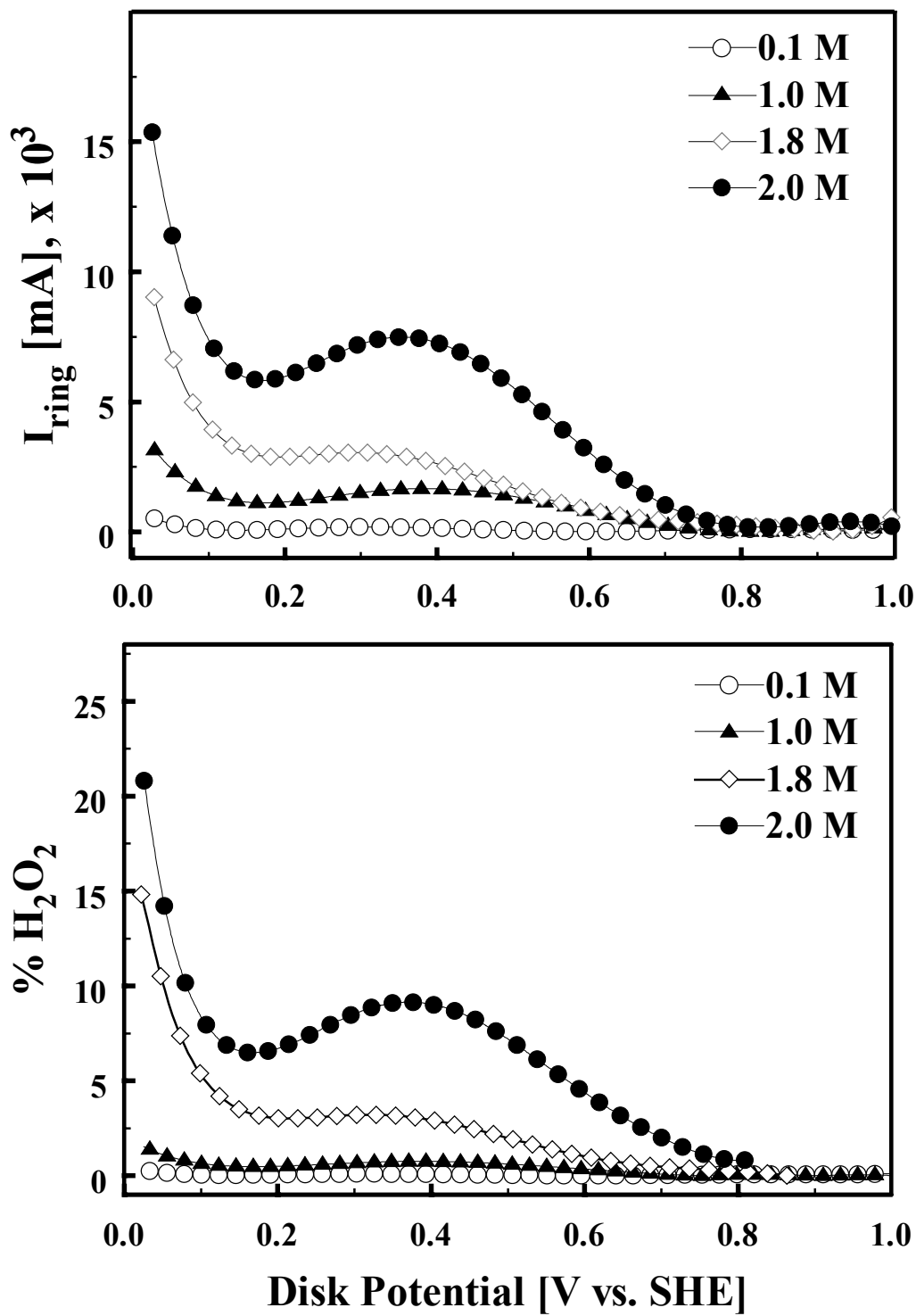

Figure 4: (a) Rate and (b) % of $H_2O_2$ formed during the oxygen reduction reaction on Pt/Vulcan XC-72R in 0.1 M [○], 1.0 M [▲], 1.8 M [□] and 2.0 M [●] $HClO_4$ solution [pH = -0.3] bubbled with $O_2$ at 25 °C. $E_{ring}$ = 1.2 V vs. SHE, 1mV/s, 2500 rpm.

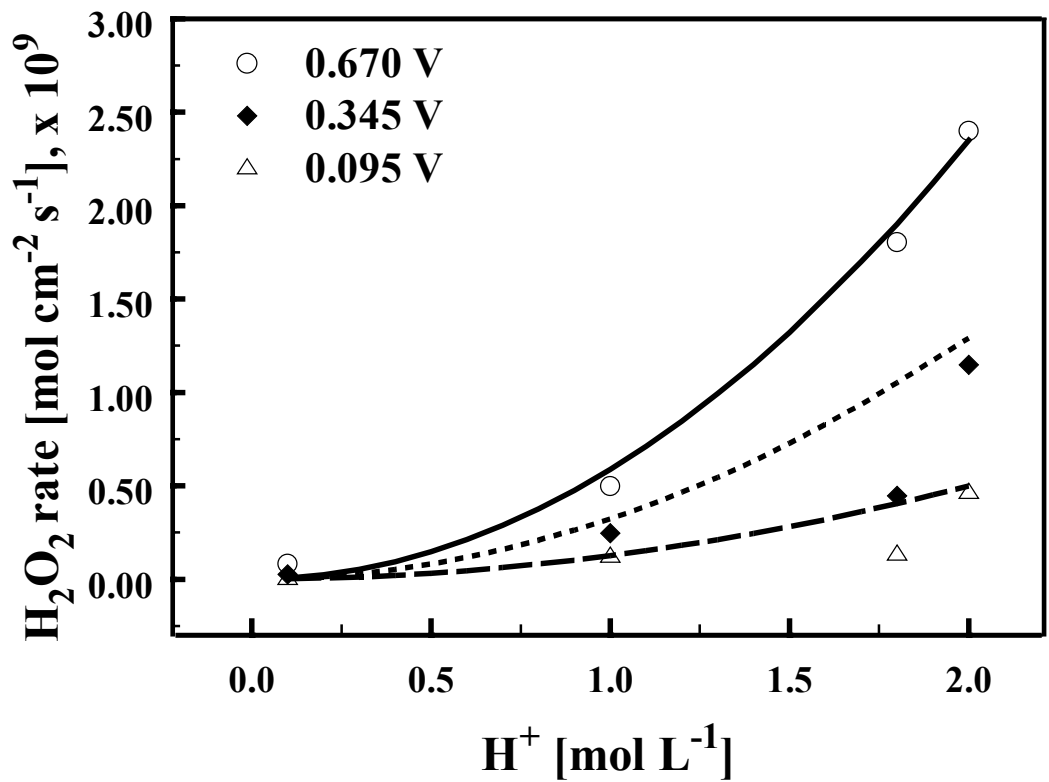

Figure 5: Mass transport corrected $H_2O_2$ formation rates [mol cm$^{-2}$ s$^{-1}$] on Pt/Vulcan XC-72R in $HClO_4$ solution as a function of acidity [M] for the following three overpotentials: 0.695 V [○], 0.345 V [□] and 0.095 V [□]. The lines are predictions according to Equation 13. The reaction order with respect to H$^+$ in the $H_2O_2$ formation reaction, $b = 2$.

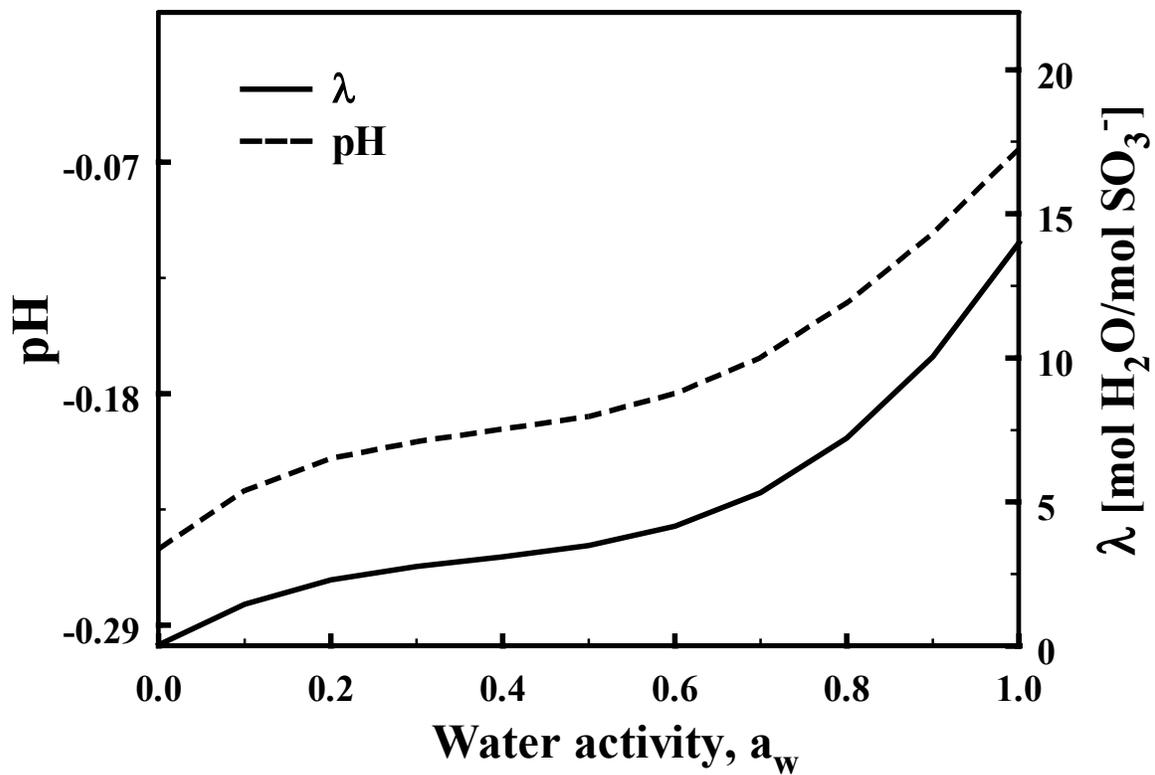

Figure 6: (a) pH (---) and λ (—) vs. water activity, $a_w$, plots for Nafion®. λ is the amount of water per sulphonic acid group [mol basis] and $a_w$ is the effective mole fraction of water and is equal to the equilibrium relative humidity expressed as a fraction.

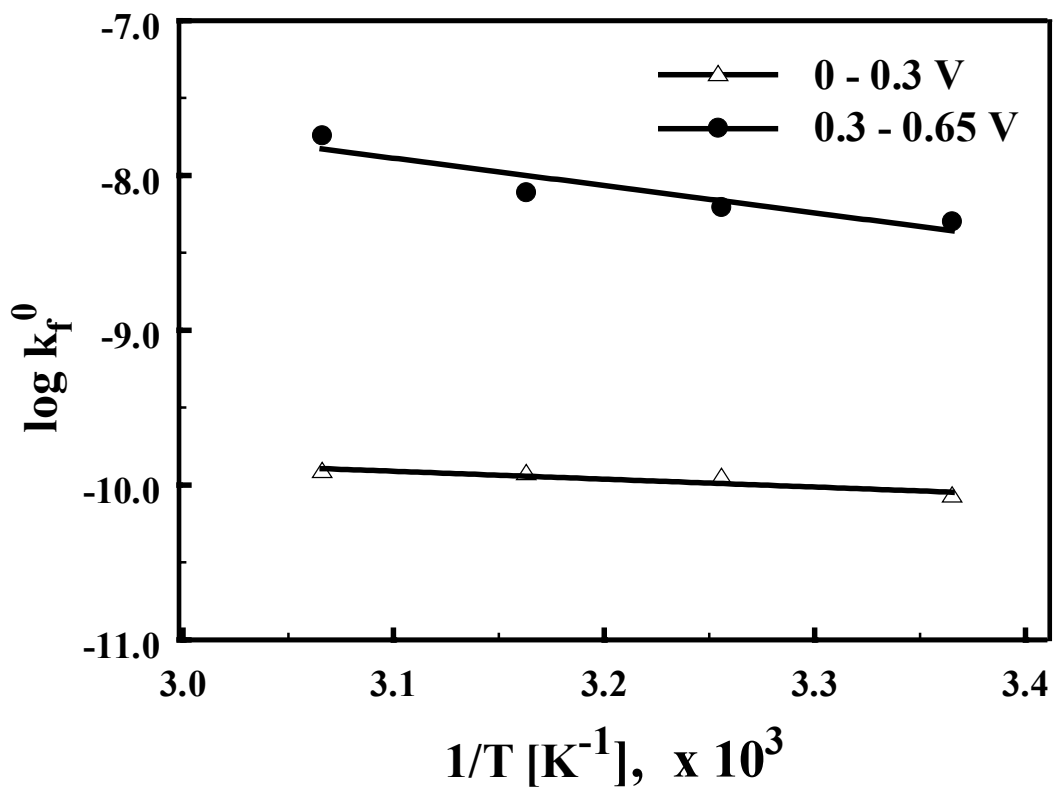

Figure 7: Arrhenius plots for $H_2O_2$ formation reaction on Pt/Vulcan XC-72R in 2M $HClO_4$ at overpotential of 0.195 V (-○-) and 0.695 V (-●-) vs. NHE.

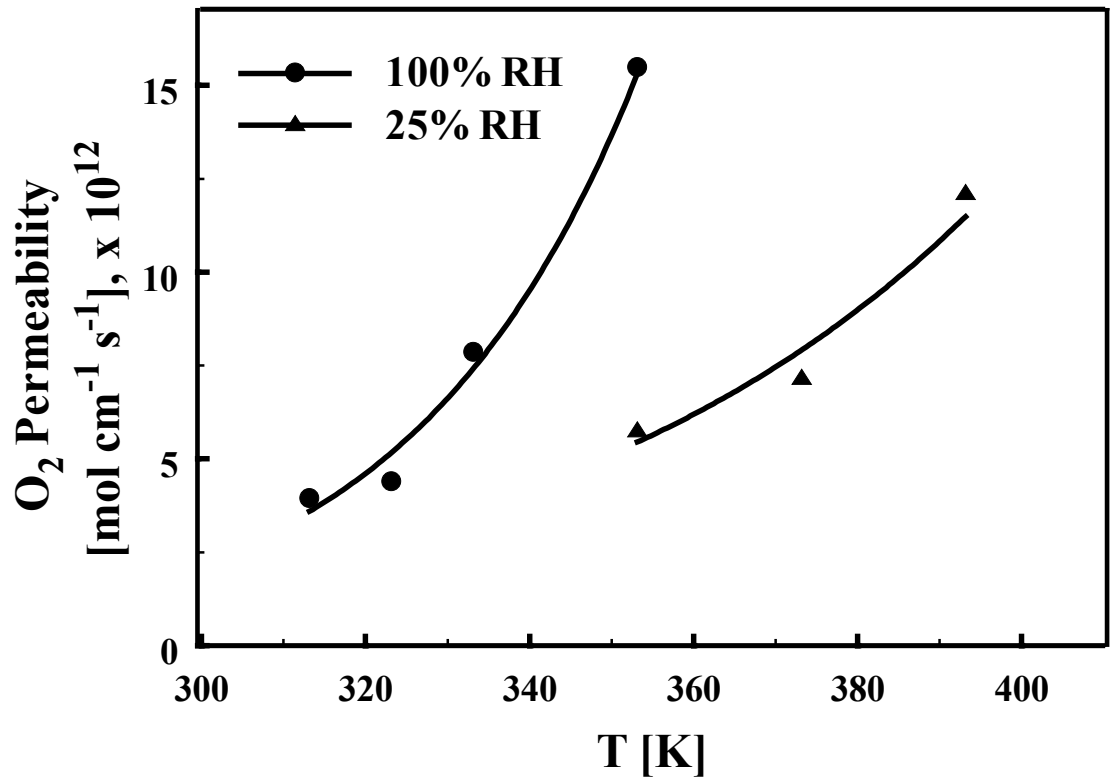

Figure 8: Oxygen permeability [mol cm$^{-1}$ s$^{-1}$] in Nafion® 112 measured by electrochemical monitoring technique as a function of temperature for two different relative humidities, 25% and 100% RH. These rates were normalized to a 101 kPa $O_2$ feed.

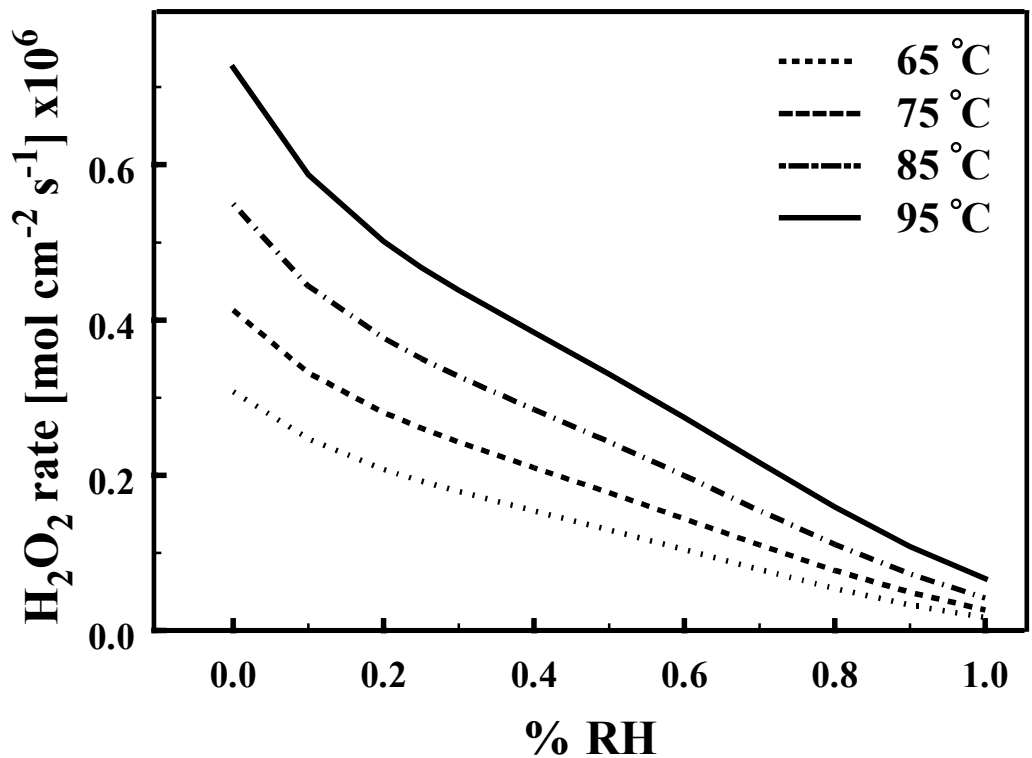

Figure 9: Rates of H$_2$O$_2$ formation [mol cm$^{-2}$ s$^{-1}$] in the cathode side of a PEM fuel cell for different relative humidities and temperatures. Local potential at the cathode was assumed to be ~0.6 V vs. SHE, which translates to an overpotential of 0.095 V for H$_2$O$_2$ formation.

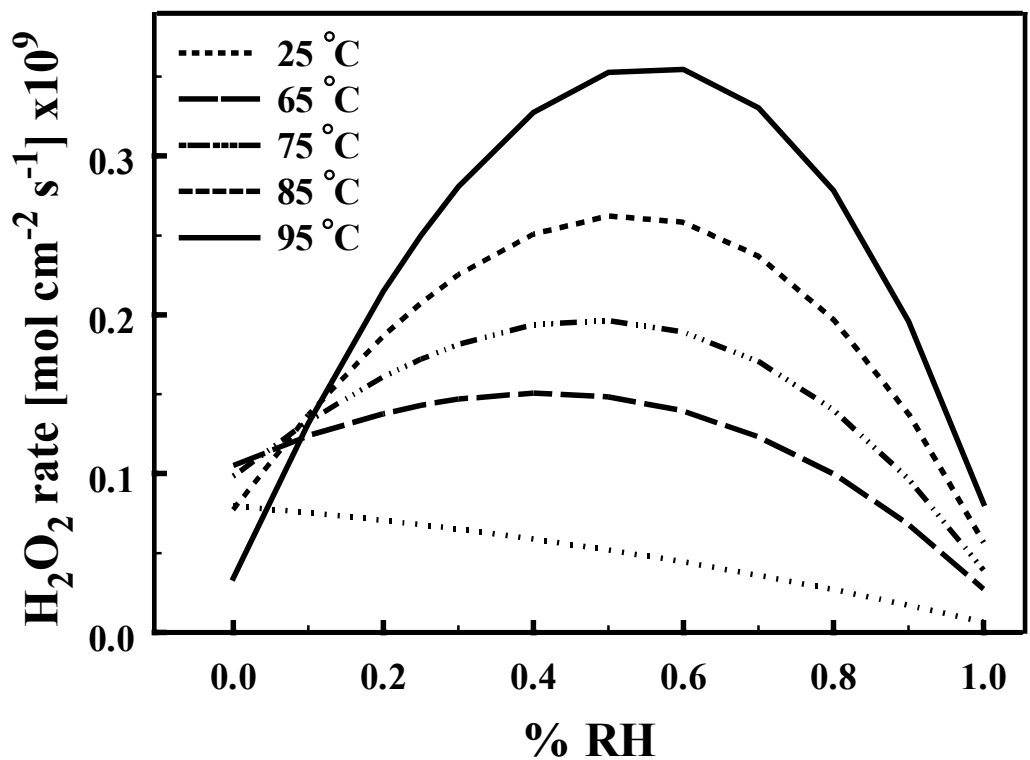

Figure 10: Rates of H$_2$O$_2$ formation [mol cm$^{-2}$ s$^{-1}$] in the anode side of a PEM fuel cell for different relative humidities and temperatures. Local potential at the anode was assumed to be ~0 V vs. SHE, which translates to an overpotential of 0.695 V for H$_2$O$_2$ formation.

# References

1. http://www.eere.energy.gov/hydrogenandfuelcells/

2. H. A. Gasteiger, W. Gu, R. Makharia, M. F. Mathias and B. Sompalli, in *Handbook of Fuel Cells – Fundamentals, Technology and Applications*, Volume 3, p. 593, W. Vielstich, A. Lamm and H. A. Gasteiger (Editors), Wiley, New York (2003).

3. D. P. Wilkinson and J. St-Pierre, *Handbook of Fuel Cells: Fundamentals, Technology and Applications*, Volume 3, p. 611, W. Vielstich, A. Lamm and H. A. Gasteiger (Editors), Wiley, New York (2003).

4. C. Paik, T. Skiba, V. Mittal, S. Motupally, and T. Jarvi, Abstract 771, *The Electrochemical Society Meeting Abstracts*, Volume 2005-1, Quebec City, Canada, May 15-20, 2005.

5. V. Mittal, C. Paik and S. Motupally, Abstract 772, *The Electrochemical Society Meeting Abstracts*, Volume 2005-1, Quebec City, Canada, May 15-20, 2005.

6. S. Hommura, K. Kawahara, and T. Shimohira, Abstract 806, *The Electrochemical Society Meeting Abstracts*, Volume 2005-1, Quebec City, Canada, May 15-20, 2005.

7. M. Inaba, H. Yamada, J. Tokunaga, and A. Tasaka, Abstract 1506, *The Electrochemical Society Meeting Abstracts*, Volume 2005-1, Quebec City, Canada, May 15-20, 2005.

8. T. Takeshita, F. Miura, and Y. Morimoto, Abstract 1511, *The Electrochemical Society Meeting Abstracts*, Volume 2005-1, Quebec City, Canada, May 15-20, 2005.

9. A. B. LaConti, M. Hamdan and R. C. McDonald, in *Handbook of Fuel Cells – Fundamentals, Technology and Applications*, Volume 3, p. 647, W. Vielstich, A. Lamm and H. A. Gasteiger (Editors), Wiley, New York (2003).

10. E. Endoh, S. Terazono, H. Widjaja and Y. Takimoto, *Electrochem. Solid State Letters,* **7**, A209 (2004).

11. A. Panchenko, *Dipl.-Chem.*, Institute für Phzsikalische Chemie der Universität Stuttgart, October 2004.

12. A. Panchenko, H. Dilger, J. Kerres, M. Hein, A. Ullrich, T. Kaz and E. Roduner, *Phys. Chem. Chem. Phys.,* 6, 2891 (2004).

13 . W. Bi, G. E. Gray and T. Fuller, *Electrochem. Solid State Letters*, **10**, B101 (2007).

14. D. E. Curtin, R. D. Lousenberg, T. J. Henry, P. C. Tangeman and M. E. Tisack, in *Proceedings of the 2$^{nd}$ European Polymer Electrolyte Fuel Cell Forum*, Vol. 1, p. 25-35 (2003); *J. Power Sources*, **131**, 41-48 [2004].

15. W. Liu and D. Zuckerbrod, *J. Electrochem. Soc.,* **152**, A1165 (2005).

16. U. A. Paulus, T. J. Schmidt, H. A. Gasteiger and R. J. Behm, *J. Electroanal. Chem.,* **495**, 134 (2001).

17. E. Claude, T. Addou, J.-M. Latour and P. Aldebert, *J. Applied Electrochemistry*, **28**, 57 (1998).

18. U. A. Paulus, A. A. Wokaun, G. G. Scherer, T. J. Schmidt, V. Stamenkovic, N. M. Markovic, and P. N. Ross, *Electrochim. Acta,* **47**, 3787 (2002).

19. O. Antoine and R. Durand, *J. Applied Electrochemistry,* **30**, 839 (2000).

20. M. A. Enayetullah, T. D. DeVilbiss, and J. O'M. Bockris, *J. Electrochem. Soc.,* **136**, 3369 (1989).

21. V. S. Murthi, R. C. Urian and S. Mukerjee, *J. Phys. Chem. B*, **108**, 11011 (2004).
22. T. J. Schmidt, H. A. Gasteiger, G. D. Stäb, P. M. Urban, D. M. Kolb and R. J. Behm, *J. Electrochem. Soc.,* **145**, 2354 (1998).
23. T. J. Schmidt, U. A. Paulus, H. A. Gasteiger and R. J. Behm, *J. Electroanal. Chem.*, **508**, 41 (2002).
24. N. Markovic and P. N. Ross, *J. Electroanal. Chem.*, 330, **499** (1992).
25. V. Stamenkovic, N. M. Markovic and P. N. Ross, Jr., *J. Electroanal. Chem.,* **500**, 44 (2001).
26. L. H. Gevantman, in *CRC Handbook of Chemistry and Physics*, 79th ed., D. R. Lide, Editor, p. 8-86, CRC Press, New York (1998-99).
27. W. J. Albery and M. L. Hitchman, *Ring-Disc Electrodes*, Clarendon Press, Oxford (1971).
28. *Rotated Ring-Disk Electrodes: DT21 Series*, Pine Instrument Company, Raleigh, NC 27617.
29. S. K. Zečević, J. S. Wainright, M. H. Litt, S. Lj. Gojković and R. F. Savinell, *J. Electrochem. Soc.,* **144**, 2973 (1997).
30. A. Damjanovic and G. Hudson, *J. Electrochem. Soc.,* **135**, 2269 (1988).
31. U. A. Paulus, A. Wokaun, G. G. Scherer, T. J. Schmidt, V. Stamenkovic, V. Radmilovic, N. M. Markovic and P. N. Ross, *J. Phys. Chem. B,* **106**, 4181 (2002).
32. A. B. Anderson and T. V. Albu, *J. Electrochem. Soc.,* **147**, 4229 (2000).
33. R. A. Sidik and A. B. Anderson, *J. Electroanal. Chem.,* **528**, 69 (2002).
34. Y. Wang and P. B. Balbuena, *J. Chem. Theory Comput.,* **1**, 935 (2005).
35. D. R. Morris and X. Sun, *J. Appl. Polym. Sci.,* **50**, 1445 (1993).
36. D. Rivin, C. E. Kendrick, P. W. Gibson, N. S. Schneider, *Polymer,* **42**, 623 (2001).
37. N. H. Jalani, P. Choi and R. Datta, *J. Membrane Sci.,* **254**, 31 (2005).
38. T. Zawodzinski, M. Neeman, L. Sillerud and S. Gottesfeld, *J. Phys. Chem.,* **95**, 6040 (1991).
39. J. T. Hinatsu, M. Mizuhata and H. Takenaka, *J. Electrochem. Soc.,* **141**, 1493 (1994).
40. S. Motupally, A. J. Becker and J. W. Weidner, *J. Electrochem. Soc.,* **147**, 3171 (2000).
41. T. Fuller, *Ph. D. Dissertation*, University of California, Berkeley (1989).
42. R. C. Reid, J. M. Prausnitz and T. K. Sherwood, *The Properties of Gases and Liquids*, McGraw Hill Inc., New York (1977).
43. K. C. Neyerlin, W. Gu, J. Jorne and H. A. Gasteiger, *J. Electrochem. Soc.,* **153**, A1955 (2006).
44. A. J. Bard and L. R. Faulkner, *Electrochemical Methods*, Wiley & Sons, Inc., New York (1980).
45. T. Sakai, H. Takenaka and E. Torikai, *J. Electrochem. Soc.,* **133**, 88 (1986).
46. K. Broka and P. Ekdunge, *J. Appl. Electrochem.,* **27**, 117 (1997).
47. S. Burlatsky, T. Jarvi, and V. Atrazhev, Abstract # 502, *The Electrochemical Society Meeting Abstracts*, Volume 2005-2, Los Angeles, California, October 16-21, 2005.
48. Q. Guo, V. A. Sethuraman, and R. E. White, *J. Electrochem. Soc.,* **151**, A983 (2004).
49. K. A. Mauritz and R. B. Moore, *Chemical Reviews,* **104**, 4535 (2004).
50. R. H. Perry and D. W. Green, Perry's Chemical Engineer's Handbook, 7th Edition, McGraw Hill (1997).